\documentstyle[12pt,preprint,aps,floats,epsfig]{revtex}
\input epsf 
\def\ni{\noindent}
\begin{document}

\title{Neutrino Pair Emission from Cooper Pair Breaking \\ and
Recombination in Superfluid Quark Matter}
\author{Prashanth Jaikumar and  Madappa Prakash } 
\address{ Department of Physics \& Astronomy, SUNY at Stony Brook, New York
          11794-3800 }
\date{\today} 
\maketitle 
\tightenlines
\begin{abstract}
For the low energy Standard Model neutrino-matter interactions, we
calculate neutrino pair ($\nu\bar\nu$) emissivites in superfluid quark
matter.  Just below the critical temperature, Cooper pairs
continuously break and recombine, resulting in the emission of
$\nu\bar \nu$ pairs with a rate that greatly exceeds the standard
quark modified Urca and bremsstrahlung rates.  At the same baryon
density in baryonic and quark matter, the ratio of baryon to quark
$\nu\bar\nu$ emissivites lies in the range $2-5$ for the densities of
interest in the long-term cooling of solar mass compact stars.  We
also find that in matter containing hyperons, $\nu\bar\nu$ emission
can occur with hyperons of all species.

\end{abstract}

In a prescient paper, Flowers, Ruderman, and Sutherland~\cite{Flo}
showed that just below the critical temperature ${T_{c}}$, the
neutrino pair ($\nu\bar\nu$) emissivity from the breaking and
recombination of superfluid neutron pairs ($n + n\rightarrow n + n +
\nu + \bar\nu$) greatly exceeds that from the so-called modified Urca
processes ($n + n \rightarrow n + p + e^- + \bar\nu$ and $n + p
\rightarrow n + n + e^+ + \nu$) in neutron star interiors.  It is only
recently that the pair breaking and formation (PBF) processes have
been included in calculations of the long term cooling of neutron
stars~\cite{SchaDany}.  The principal effect of the PBF process is to
rapidly cool the star during its early hundreds of years of thermal
evolution, until competing $\nu$-emitting processes (for recent
reviews see Ref.~\cite{Yaks}) begin to dominate the further cooling of
the star for several thousands or million years that they might remain
observable with such instruments as HST, Chandra, and XMM.  The
precise extent to which the core temperature falls depends on the
smallest of the pairing gaps among the superfluid species present in
the star~(cf. \cite{ppls} and references therein). To date, the role
of PBF processes has been explored only for baryon
superfluidity~\cite{SchaDany}.

In this work, we calculate the $\nu\bar\nu$ emissivity in superfluid
quark matter likely to be encountered in neutron star
interiors~\cite{SFs}.  For the low energy Standard Model
neutrino-matter interactions, we show that the predominant
contribution to the total $\nu\bar\nu$ emissivity comes from the
vector channel, while the axial channel provides a small correction.
This heirarchy is similar to that found in baryonic matter~\cite{Flo}.
We also consider the effect of quark masses on the emissivities in
limiting cases. We compare our results for $\nu\bar \nu$
emissivities with those of the modified Urca and bremsstrahlung
processes in both quark and nucleonic matter~\cite{Iwas,FM,HPR}.
Expected $\nu\bar\nu$ emissivities from hyperons in baryonic matter
and from $p$-wave pairing in quark matter are also addressed. 
\clearpage

The Hamiltonian that describes the neutral current quark-neutrino
weak interaction is given by~\cite{Wein}
\begin{equation}
H_{W} =
\frac{G_{F}}{\sqrt{2}}~ [{\bar{\psi}}^{i,a}\gamma^{\lambda}(c_{V} -
c_{A}\gamma_{5})\psi_{i,a}]~[{\bar{\psi}}_{\nu_{\ell}}\gamma_{\lambda}(1 -
\gamma_{5})\psi_{\nu_{\ell}}] \,,
\end{equation}
here $G_{F}$ is the Fermi constant for the weak interaction, and
$c_{V}$ and $c_{A}$ are flavor dependent vector and axial-vector
coupling constants, respectively.  The index ${i}$ denotes quark
flavor (${N_{f} = 3}$) and ${a}$ denotes color (${N_{c} = 3}$).  
The index $\ell$ stands for lepton flavor $e,\mu$ or $\tau$. At
leading order (tree level), flavor changing processes do not occur
through the weak interaction, hence the diagonal structure in flavor
and color space.  Unlike baryons, quarks in a superfluid phase can lock
themselves in preferred directions in color and flavor
space~\cite{SFs}.  However, we are considering neutrino pair
($\nu\bar\nu$) production mediated by the neutral current weak interaction, which is blind to color and flavor.  These degrees of freedom are, however,
counted in the density of states at the Fermi surface.

We begin with the case in which quarks of all flavors are taken to be
massless.  In this case, the full structure of the weak interaction
through 4-component Weyl spinors is needed (In the nonrelativistic
case, {\it e.g.} for baryons in the superfluid phase, the V--A
interaction has large matrix elements only in the 2${\times}$2 block
diagonals in the space of 4-component spinors.  This allows for
simpflication in that 2-component spinors with a suitable
normalization are adequate.) The helicity basis (which is also the
chirality basis for massless particles when we identify positive
helicity with right handed particles $\{R\}$ and negative helicity
with left handed $\{L\}$ ones) is convenient in this case.  The quark
field operator may be expressed as~\footnote {In this form, the quark
field operator does not take into account the fact that diquarks may
condense in preferred channels in color and flavor space, as, for
instance in the three flavor color-flavor locked (CFL) phase, or the
two flavor superconducting (2SC) phase~(see, for example,
\cite{CFL}). We return later to comment on the expected modifications
in ``QCD condensed matter''.}

\begin{equation}
\psi(\vec{\bf{x}}) =
\sum_{\bf{p}\rm} \biggl(e^{-i\vec{\bf{p}}.\vec{\bf{x}}}
u_{p}\{\chi_{\vec{\bf{p}},R}b_{\vec{\bf{p}},R}+
\chi_{\vec{\bf{p}},L}b_{\vec{\bf{p}},L}\} +
e^{i\vec{\bf{p}}.\vec{\bf{x}}}
v_{p}\{\chi_{\vec{\bf{p}},R}{b}^{\dag}_{\vec{\bf{p}},L} -
\chi_{\vec{\bf{p}},L}{b}^{\dag}_{\vec{\bf{p}},R}\}\biggr) \,,
\label{fop}
\end{equation}
where $\chi$'s are 4-component Weyl spinors and we have paired
particles of opposite helicity according to
\begin{eqnarray}
b_{\vec{\bf{p}},R} = u_{p} a_{\vec{\bf{p}},R} + v_{p}
{a}^{\dag}_{-\vec{\bf{p}},L} \qquad  {\rm and} \qquad 
b_{\vec{\bf{p}},L} = u_{p}
a_{\vec{\bf{p}},L} - v_{p} {a}^{\dag}_{-\vec{\bf{p}},R} \,.
\end{eqnarray}
The quasiparticle excitations are determined through a Bogoliubov-Valatin
transformation on the usual creation and annihilation
operators ($a$ and ${a}^\dag$), and are distributed according to the
Fermi-Dirac function at finite temperature.  For isotropic $^1{\rm S}_0$
pairing, the coherence factors ${u_{p}}$ and ${v_{p}}$ satisfy
\begin{equation}
u_{p}v_{p} = \frac{\Delta}{2E_{p}},\qquad u_{p}^{2} =
\frac{1}{2}\biggl(1+\frac{\epsilon_{p}}{E_{p}}\biggr),\qquad {\rm and} \qquad
 v_{p}^{2} =
\frac{1}{2}\biggl(1-\frac{\epsilon_{p}}{E_{p}}\biggr) \,\label{coh} .
\end{equation}
The quasiparticle excitations carry momentum ${\bf{\it{p}}}$ and
energy $E_{p} = \sqrt{ {\epsilon_{p}}^{2} + {\Delta}^{2} }$, where
$\epsilon_{p} = p-\mu$ and $\Delta$ is the superfluid gap.  The field
operator in Eq.~(\ref{fop}) acts on a quasiparticle state
${|p_{R},p^{\prime}_{L}\rangle}$ and restores it to the
condensate. Conversely, it can also excite the condensate so that pair
breaking and recombination occur simultaneously.  We denote the matrix
element for this process by ${M_{q}}$ and the neutrino matrix element
by ${M_{\nu}}$. The time dependence on the field operator has been
dropped, since the quasiparticle lifetime is essentially infinite
compared to the weak interaction time. Then, the expression for the
emissivity is
\begin{eqnarray}
\epsilon  =
         \frac{1}{2}N_\nu N_{f}N_{c}\biggl(\frac{G_{F}}{\sqrt{2}}\biggr)^{2}
\frac{(2\pi)^{4}}{(2\pi)^{12}} \int
         \hspace{0.05in} \frac{d^{3}p}{2p_{0}} &&
         \frac{d^{3}p^{\prime}}{2p_{0}^{\prime}}\,f(E_{p})f(E_{p^{\prime}}) 
	\nonumber \\  
	 && \int  \hspace{0.05in} \frac{d^{3}q_{1}}{2q_{10}}
         \frac{d^{3}q_{2}}{2q_{20}}~
	(q_{10} +q_{20})
	|M_{q}|^{2}|M_{\nu}|^{2} ~\delta^{4}(p+p^\prime-q_1-q_2) \,,
\label{emis}
\end{eqnarray}
where $p$ and $p^\prime$ denote the four-momenta of quarks and $q_1$
and $q_2$ those of the neutrinos. The number of neutrino flavors is
denoted by $N_\nu$.  The simple factor of ${N_{f}N_{c}}$ outside
arises, because we have disregarded strong interaction effects and
flavor changing processes that occur at higher orders. The overall
factor of 1/2 avoids double counting of the same quasiparticle
collision.  $f$ denotes the Fermi-Dirac distribution function.
Summing over the neutrino spins yields the neutrino squared matrix
element, $|M_{\nu}|^{2} = 8
(q_{2\mu}q_{1\nu}+q_{1\mu}q_{2\nu}-g_{\mu\nu}q_{1}.q_{2}
-i\epsilon_{\mu\nu\alpha\beta}q_{1\alpha}q_{2\beta})$.  The quark
matrix element receives contributions from the vector ($V$), axial
(${A}$) and the mixed vector-axial ($M$) channels. \\ The vector part
of the interaction acting on the state
${|p_{R},p^{\prime}_{L}\rangle}$ gives
\begin{equation}
(|M_{q}|^{2})^{\mu\nu}_{V} =
c_{V}^{2} (u_{p}v_{p^{\prime}}+v_{p}u_{p^{\prime}})^{2}
({\bar{\chi}}_{\vec{\bf{p}},L}
\gamma^{\mu}{\chi}_{\vec{\bf{p^{\prime}}},L}
{\bar{\chi}}_{\vec{\bf{p^{\prime}}},L}\gamma^{\nu}{\chi}_{\vec{\bf{p}},L}) \,,
\end{equation}
where we have used the spinor identity 
$\bar{\chi}_{\vec{\bf{p}},R}\gamma^{\alpha}{\chi}_{\vec{\bf{p^{\prime}}},R}
=
\bar{\chi}_{\vec{\bf{p^{\prime}}},L}
\gamma^{\alpha}{\chi}_{\vec{\bf{p}},L} ~\cite{Itz}$ in writing the term in 
the rightmost bracket. 
Using the standard normalization for massless spinors, the
vector contribution to ${|M_{q}|^{2}}$ becomes
\begin{equation}
(|M_{q}|^{2})^{\mu\nu}_{V} =
2(u_{p}v_{p^{\prime}}+v_{p}u_{p^{\prime}})^{2} \cdot
\frac{c_V^2}{4} {\cal T}\,, 
\end{equation}
where ${\cal T} = {\rm Tr}\left[p{\hskip-2.0mm}/\gamma^{\mu}
(1-\gamma_{5})p{\hskip-2.0mm}/^{\prime}\gamma^{\nu}(1-\gamma_{5})\right]$. 
Utilizing the relation 
$\bar{\chi}_{\vec{\bf{p^{\prime}}},R}\gamma^{\alpha}
\gamma_{5}{\chi}_{\vec{\bf{p}},R}
= -\bar{\chi}_{\vec{\bf{p}},L}\gamma^{\alpha}
\gamma_{5}{\chi}_{\vec{\bf{p^{\prime}}},L}
=
\bar{\chi}_{\vec{\bf{p}},L}\gamma^{\alpha}{\chi}_{\vec{\bf{p^{\prime}}},L}$, 
the squared matrix elements for the axial and mixed contributions are found to
be 
\begin{eqnarray}
(|M_{q}|^{2})^{\mu\nu}_{A} =
2(u_{p}v_{p^{\prime}}-v_{p}u_{p^{\prime}})^{2} \cdot \frac{c_A^2}{4}
{\cal T} \quad {\rm and}\quad 
(|M_{q}|^{2})^{\mu\nu}_{M} =
2({u_{p^{\prime}}^{2}v_{p}}^{2}-{v_{p^{\prime}}^{2}u_{p}}^{2})
\cdot\frac{2c_Vc_A}{4} {\cal T} \,.
\end{eqnarray}
The total contribution from the quark squared matrix element is simply
the sum  \\ $|M_{q}|^{2} = |M_{q}|^{2}_{V} + |M_{q}|^{2}_{A} +
|M_{q}|^{2}_{M}$. The contraction of the
neutrino and quark squared matrix elements yields, 
up to differences in couplings and coherence
factors, a generic expression of the form
\begin{equation}
|M_{q}|^{2}_{C}\times |M_{\nu}|^{2} \propto 
[16(q_{1}\cdot p)(q_{2} \cdot p^{\prime}) + 
48(q_{2}\cdot p)(q_{1} \cdot p^{\prime})] \,,
\end{equation}
where $C$ denotes any of the channels $V, A$, and $M$.

We turn now to the phase space integrations.  The neutrino phase space
integral is of the form 
\begin{eqnarray} 
N_{\alpha\beta} &=& \int
\hspace{0.05in} \frac{d^{3}q_{1}}{2q_{10}} \frac{d^{3}q_{2}}{2q_{20}}~
(q_{1\alpha}q_{2\beta}+q_{2\alpha}q_{1\beta}) (q_{10} + q_{20})
\hspace{0.05in} \delta^{4}(q_{1} + q_{2}-Q) \nonumber \\ &=& \frac{\pi
Q_{0}}{12}(Q^{2}g_{\alpha\beta} +
2Q_{\alpha}Q_{\beta})\Theta(Q_{0})\Theta(Q_{0}^{2} -
|\bf{\vec{Q}}|^{2}) \,,
\end{eqnarray}
where an additional delta function in the total neutrino energy
${Q_{0}}$ and momentum ${\vec{\bf{Q}}}$ has been introduced.  The
obvious symmetry between $\alpha$ and $\beta$ in the above integral
may be exploited to simplify the calculation ({\it e.g.} the
${\epsilon_{\mu\nu\alpha\beta}}$ terms from the neutrino squared
matrix element vanish once the neutrino integral is performed).  The
momentum conserving delta function in Eq.~(\ref{emis}) can be used to
eliminate the integral over ${{\vec{\bf{p}}}^{\prime}}$.  Next, we
shift the ${\vec{\bf{p}}}$ integral using ${\vec{\bf{p}}\rightarrow
\vec{\bf{p}}+ \frac{\vec{\bf{Q}}}{2}}$, and in view of the restriction
that ${|\vec{\bf{Q}}| \leq Q_{0}\sim \Delta}$ and ${\Delta \ll |
\vec{\bf{p}}|\sim \mu}$, the energy conserving delta function becomes
${\delta(Q_{0}-2E_{p})}$.  In addition, the two Fermi-Dirac functions
both become ${f(E_{p})}$.~\footnote{It is possible to calculate
corrections resulting from $E_p \neq E_{p^\prime}$, but lead to more
complicated expressions than found here ({\it e.g}, see
Ref.~\cite{kam,ykl}).}

In the vector channel, the momentum shift
${\vec{\bf{p}}\rightarrow \vec{\bf{p}}+ \frac{\vec{\bf{Q}}}{2}}$ allows us 
to write
\begin{equation}(u_{p}v_{Q-p}+v_{p}u_{Q-p})^{2}
\stackrel{\vec{\bf{p}}\rightarrow \vec{\bf{p}} + 
\frac{\vec{\bf{Q}}}{2}} 
{\longrightarrow}u^2_{p+\frac{Q}{2}}v^2_{-p+\frac{Q}{2}}
+v^2_{p+\frac{Q}{2}}u^2_{-p+\frac{Q}{2}} + 
2u_{p+\frac{Q}{2}}v_{p+\frac{Q}{2}}v_{-p+\frac{Q}{2}}u_{-p+\frac{Q}{2}} \,.
\end{equation}
Using Eq.~(\ref{coh}) and 
the expression for the quasiparticle energy, the first two terms become
\begin{equation}
\biggl(\frac{1}{4}\frac{\Delta^{2}}{E_{p}^{2}} +
\frac{x}{2}\frac{\Delta^{2}}{E_{p}^{2}}\frac{\epsilon_{p}}{E_{p}} +
(x\rightarrow -x)\biggr) + O(x^{2}) \,,
\end{equation}
where ${x = {\Delta}/{\mu} \ll 1}$, justifying the neglect of the
${O(x^{2})}$ terms. The third term becomes $\Delta^2/(2E_p^2)$.  Thus,
the coherence factors for the vector channel take the form
\begin{equation}
(u_{p}v_{p^{\prime}}+v_{p}u_{p^{\prime}})^{2} 
 = \frac{\Delta^2}{E_p^2} \,. \label{vcoh}\qquad
\end{equation}
For the axial contribution to the quark matrix element, 
we retain terms up to ${O(x^{2})}$ to get
\begin{eqnarray}
(u_{p}v_{Q-p}-v_{p}u_{Q-p})^{2}\stackrel{\vec{\bf{p}}\rightarrow
\vec{\bf{p}}+ \frac{\vec{\bf{Q}}}{2}}{\longrightarrow} =
\frac{(\vec{\bf{Q}}.\hat{\bf{p}})^{2}{\Delta}^{2}}{4E_{p}^{4}} \,.
\label{acoh}
\end{eqnarray}
Eqs.~(\ref{vcoh}) and ~(\ref{acoh}) enable the 
coherence factor for the mixed contribution to be written as 
\begin{equation}
({u_{p^{\prime}}^{2}v_{p}}^{2}-{v_{p^{\prime}}^{2}u_{p}}^{2}) =
\sqrt{\frac{\Delta^{2}}{E_{p}^{2}}\frac{(\vec{\bf{Q}}.
\hat{\bf{p}})^{2}{\Delta}^{2}}{4E_{p}^{4}}}
= \frac{\vec{\bf{Q}}.\bf{\hat{p}}\rm\Delta^{2}}{2E_{p}^{3}} \,.
\end{equation}
The mixed contribution, being odd in ${\bf{\vec{Q}}}$, 
vanishes on angular integration.
Integrals over $\vec{\bf{Q}}$ can be performed using the formula 
\begin{equation}
\int
d\Omega_{q}  d|\vec{\bf{Q}}|~|\vec{\bf{Q}}|^{m}~ 
(|\vec{\bf{Q}}\cdot{\bf{\vec{p}}}|)^{
n} = \frac{4\pi}{n+1}|\vec{\bf{p}}|^{n}\int
d|\vec{\bf{Q}}|~|\vec{\bf{Q}}|^{m+n} \,.
\end{equation}
The integrals over ${Q_{0},\vec{\bf{Q}}}$, and ${d\Omega_{p}}$ are now
straightforward, leaving only the integral over ${|\vec{\bf{p}}|}$. We
can make a change of variable at the Fermi surface from
$|\vec{\bf{p}}|$ to $E_{p}$ to cast the final result as a
1-dimensional integration over $E_{p}$.  Collecting these results and
summing the emissivities in the vector (V) and axial (A) channels, the
$\nu\bar\nu$ emissivity from PBF in matter comprised of massless
quarks is
\begin{eqnarray}
\epsilon^R(0) = 
N_\nu 
\left(\frac{G_{F}c_{V}}{\sqrt{2}}\right)^{2} \nu(0)(k_BT)^{7}
\left(\frac{32}{15\pi^{3}}F(y)\right)~\left[1 + 
\frac {1}{7}\left(\frac{c_A}{c_V}\right)^2\right] \,,
\label{epsr}
\end{eqnarray}
where $\nu(0) = N_fN_c\mu^{2}/\pi^{2}$ is the density of states 
at the Fermi surface $\mu$. 
The explicit form of the function $F(y)$ is
\begin{equation}
F(y) =
y^{2}\int_{y}^{\infty}\hspace{0.05in}dx
\frac{x^{5}}{\sqrt{x^{2}-y^{2}}}\frac{1}{(e^{x}+1)^{2}} \label{F} \,,
\quad
\end{equation}
where $y = \beta\Delta = \Delta/k_BT$ and $x = E_{p}/k_BT$. 
The gap $\Delta$ depends on the temperature. In limiting situations, 
the temperature dependence of the superfluid gap $\Delta$ 
for $^1{\rm S}_0$ pairing is given by
\begin{eqnarray}
\Delta(T) &=& \Delta (0) - (2\pi
 k_BT\Delta (0))^{1/2}\mbox{exp}\left(-\frac{\Delta (0)}{k_BT}\right) 
\hspace{5mm} (T
 \ll T_{c}) \nonumber \\ 
&=& 3.06~k_BT_{c}~(1-T/T_{c})^{1/2} \hspace{25mm} (T_{c}-T \ll T_{c}) 
\end{eqnarray}
with ${\Delta (0) = 1.76k_BT_{c}}$. In lieu of this, $F(y) \sim
(1-\frac{T}{T_{c}})$ as $T\rightarrow T^{-}_{c}$ so that it vanishes
quadratically with $\Delta$.  For $T \ll T_{c}$, $F(y)$ vanishes
exponentially as $\mbox{e}^{-2\Delta/T}$.  Accurate fits of $F(y)$
valid over the entire range of $y$ may be found in Ref.~\cite{Yaks}.

It is instructive to compare the result  Eq.~(\ref{epsr}) with
that of Flowers {\it et al.}~\cite{Flo} for nonrelativistic neutrons:
\begin{eqnarray}
\epsilon &=& N_\nu \left(\frac{G_Fc_V}{\sqrt 2}\right)^2 
\nu (0)(k_BT)^7 
\left(\frac{32}{15\pi^3}
F(y)\right)~\left[1 + 
\frac{11}{42} \left(\frac {c_A}{c_V}\right)^2 
\left(\frac{v_{F_n}}{c}\right)^2 \right]\,,
\label{epsnr}
\end{eqnarray}
where $\nu(0) = m^*p_{F_n}/\pi^2$ is the density of states, $m^*$ is
the Landau effective mass~\footnote{Defined at the Fermi surface, $m^*$ 
accounts for 
momentum-dependent interactions in a potential model description. 
In a field-theoretical description, $m^*$ accounts for scalar interactions 
that off-set the bare mass.}, and $v_{F_n}=p_{F_n}/m^*$ is the velocity at the
Fermi surface.  Since the PBF process occurs in the vicinity of the
Fermi surface, the formal similarity of the results between the
nonrelativistic and extreme relativistic cases is not surprising.  The neutral
current couplings (see Table~1), the density of states at the Fermi
surface, and the results of phase space integrations are, however,
somewhat different between the two cases.  For both relativistic quarks and
nonrelativistic neutrons, the dominating contribution arises from the
vector channel although the suppression of the axial  channel is 
more severe in the nonrelativistic case. 

The main approximation made in the emissivity calculation for massless
quarks is that $\Delta \ll \mu$. Modifications due to a finite quark
mass $m$ enter in the dispersion relation ${\epsilon_{p} =
\sqrt{p^{2}+m^{2}}-\sqrt{p_{F}^{2}+m^{2}}}$, with a corresponding
density of states $\nu(0)=N_fN_cp_F{\sqrt{p_F^2+m^2}}/\pi^2 $, and
from corrections to the quark matrix element through the spinor
normalization as $\Sigma_s u^{s}(p)\bar{u}^{s}(p) =
p{\hskip-2.0mm}/+m$.  In what follows, we set the spin
along the direction of the momentum of the massive quark.
The quark field operator in the spin basis may be written as
\begin{equation}
\psi(\vec{\bf{x}}) =
\sum_{\bf{p}\rm}\biggl(e^{-i\vec{\bf{p}}.\vec{\bf{x}}}u_{p}
\{\chi_{\vec{\bf{p}},+}b_{\vec{\bf{p}},+} + 
\chi_{\vec{\bf{p}},-}b_{\vec{\bf{p}},-}\} +
e^{i\vec{\bf{p}}.\vec{\bf{x}}}
v_{p}\{\chi_{\vec{\bf{p}},+}{b}^{\dag}_{\vec{\bf{p}},+} - 
\chi_{\vec{\bf{p}},-}{b}^{\dag}_{\vec{\bf{p}},-}\}\biggr) \,.
\label{qfop}
\end{equation}
This acts on a quasiparticle state 
${|p_{+},p^{\prime}_{-}\rangle}$ and restores it to the condensate.

The vector channel squared matrix element is
\begin{eqnarray}
(|M_{q}|^{2})^{\mu\nu}_{V} &=& 2(u_{p}v_{p^{\prime}}+v_{p}u_{p^{\prime}})^{2}~
\cdot~c_{V}^{2} {\rm Tr} 
[M_{-}(p\prime)\gamma^{\mu}M_{-}(p)\gamma^{\nu}] \,, \nonumber \\
M_{-}(p) &=& u_{-}(p)\bar{u}_{-}(p) = \frac{p{\hskip-2.0mm}/ +
m}{2}\biggl(1-\gamma_{5} n{\hskip-2.0mm}/(p)\biggr) \label{proj}
\end{eqnarray}
with $n(q) =
\biggl(\frac{|\bf{\vec{q}}|}{m},\frac{q^{0}}{m}\hat{\bf{q}}\biggr)$. 
Setting ${m=0}$ reproduces the chirality projector that
was used for the calculation in the massless case. Taking the
nonrelativistic limit ${m \gg |\bf{\vec{p}}|}$, we find that
only ${\mu = 0 = \nu}$ contributes to the matrix element. This is
consistent with the nonrelativistic matrix elements considered in the
neutron superfluid case ~\cite{Flo}.

The squared matrix element in the axial  channel is
\begin{eqnarray}
&(|M_{q}|^{2})^{\mu\nu}_{A} = 2(u_{p}v_{p^{\prime}}-v_{p}u_{p^{\prime}})^{2}~
\cdot~c_{A}^{2} {\rm Tr} 
[M_{-}(p\prime)\gamma^{\mu}\gamma_{5}M_{-}(p)\gamma^{\nu}\gamma_{5}] \,.
\end{eqnarray}
To perform the trace, notice that there are ${{\cal O}(m^2), {\cal
O}(m)}$, and ${{\cal O}(m^{0})}$ terms that may be grouped
separately. The ${{\cal O}(m)}$ terms in the vector and axial channels
vanish, because they involve the trace of an odd number of
$\gamma$-matrices or are anti-symmetric in the very indices in which
the overall neutrino integral is symmetric.

In the vector channel, the ${{\cal O}(m^{2})}$ term is
\begin{equation}
\frac{m^{2}}{4}\{ 4g^{\mu\nu} + 4n^{\mu}n^{\prime\nu} +
4n^{\nu}n^{\prime\mu} - 4(n.n^{\prime})g^{\mu\nu}\} \,,
\label{omsq}
\end{equation}
while the ${{\cal O}(m^{0})}$ term is 
\begin{equation}
\frac{1}{4}\{4p^{\mu}p^{\prime\nu} + 4p^{\nu}p^{\prime\mu} -
4(p.p^{\prime})g^{\mu\nu}\} - \frac{1}{4} 
{\rm Tr} 
[p{\hskip -2.0mm}/n{\hskip
-2.0mm}/\gamma^{\mu}p{\hskip -2.0mm}/^{\prime}n {\hskip
-2.0mm}/ ^{\prime}\gamma^{\nu}] \label{om}\,.
\end{equation}
The axial channel gives similar contributions as above, but with the
following changes: (i) the first and last terms in Eq.~(\ref{omsq})
are both of opposite signs, and (ii) the last term in Eq.~(\ref{om})
is of opposite sign. 

Simple expressions for the emissivities that take into account quark
masses may be obtained by retaining leading terms in some limiting
situations. \\

\ni (a) For $m/p_F \ll 1$, the projector in Eq.~(\ref{proj}) reads
$\frac{p{\hskip-1.5mm}/}{2}(1+\gamma_{5})+\frac{m}{2}$. 
In this case, the total emissivity is 
\begin{equation}
\epsilon^{R}(m) = \epsilon^{R}(0) 
~\left[\left(1-\frac{m^2}{4p_{F}^2}\right)+
\frac {1}{7}\left(\frac{c_A}{c_V}\right)^2
\left(1+\frac{7m^2}{12p_{F}^2}\right)\right] \,,  
\label{epsrm}
\end{equation}
where $\epsilon^{R}(0)$ is the result in Eq.~(\ref{epsr}).  In the
context of the long-term cooling of neutron stars with quark cores,
the utility of this result lies in the extent to which it applies
for massive strange quarks. \\

\ni (b) For $p_F/m \ll 1$, the projector in Eq.~(\ref{proj}) becomes 
$m(1+\gamma_{0})(1+\gamma_{5}{\bf \vec{\gamma}\cdot\hat{p}})-|{\bf
\vec{p}}|({\bf \vec{\gamma}\cdot\hat{p}}-\gamma_{0}\gamma_{5})$. \\ 
In this case,
\begin{eqnarray}
\epsilon_{V}(m) &=& \epsilon_V(0)
\left[\frac{m^2}{m^2+p_{F}^{2}}-\frac{31}{84}
\frac{p_{F}^{2}}{m^2+p_{F}^{2}}\right] \qquad {\rm and} \qquad \nonumber \\ 
\epsilon_{A}(m) &=& \left(\frac{c_{A}}{c_{V}}\right)^{2}\epsilon_{V}(0) 
\left[\left(\frac{v_{F}}{c}\right)^2
\frac{11}{42}\left(\frac{m^2+p_{F}^{2}/2}{m^2+p_{F}^{2}}\right)+
\left(\frac{p_{F}^{2}}{m^2+p_{F}^{2}}\right)\right] \,,
\label{epsnrmva}
\end{eqnarray}
where $\nu(0)$ and $F(y)$ in $\epsilon_{V}(0)$ involve the dispersion
relation for massive particles.  The changes from the extreme
nonrelativistic result in Eq.~(\ref{epsnr}) are easily understood. The
factor $m^{2}+p_{F}^{2}$ in the denominator of the two terms inside
the square bracket comes from energy factors in the Lorentz invariant
phase space, while the numerator receives contributions from
${\cal{O}}(m^{2})$ and ${\cal{O}}(m^{0})$ terms in the trace. The
heirarchy $\Delta \ll p_{F} \ll m$ should be maintained for these
expressions to be valid. We can express the sum of the emissivities
from the vector and axial channels in Eq.~(\ref{epsnrmva}) as
\begin{equation}
\epsilon^{NR}(m) = \epsilon_V(0)\left[1+
\left(\frac{c_{A}}{c_{V}}\right)^{2}\left(\frac{v_{F}}{c}\right)^2
\left\{\frac{11}{42}+
\left(\frac{m^*}{m}\right)^2 \right\}+ ... \right] \label{emnrsum}
\end{equation}
In this form, we recognize the nonrelativistic result of Flowers {\it
et al.} in Eq.~(\ref{epsnr}), and in the axial channel 
an additional term involving $(m^*/m)^2$, where $m^*$ is
the Landau effective mass (see footnote 2)
defined through $p_{F} = m^{*}v_{F}/c =
(\pi^{2}n_{B})^{1/3}$.
The ellipsis denotes terms that are smaller by further powers
of $(p_F/m)^2$ compared to the preceding term in
Eq.~(\ref{emnrsum}).  The presence of the $(m^*/m)^2$ term and its
importance in the case of superfluid protons for which $c_V$ is
vanishingly small, was first noted in Ref.~\cite{kam} and was obtained
as a relativistic correction in PBF studies of a proton
superfluid~\cite{kam}.  For further discussions on additional
corrections, see Ref.~\cite{ykl}. 

The results in Eqs.~(\ref{epsr}) and (\ref{epsrm}) are
the prinicipal new results of this work. For simplicity and for numerical
comparison with emissivities from other processes in quark and
baryonic matter, we will use the result in Eq.~(\ref{epsr}) in what follows  
to represent the $\nu\bar\nu$ emissivity from the PBF process in
quark matter.  In cgs units, Eq.~(\ref{epsr}) may be expressed as
\begin{eqnarray}
\epsilon_q^{\nu\bar\nu} 
&\cong &1.4 \times 10^{20}~N_\nu T_9^7 F~ 
a_q \left(\frac{n_B}{n_0}\right)^{2/3}~{\rm erg~cm^{-3}~s^{-1}} \,, 
\label{epsq}
\end{eqnarray}
where $N_\nu$ is the number of neutrino flavors, $T_9$ is the
temperature in ${10^9}~^{\rm o}{\rm K}$, $F$ is the result in
Eq.~(\ref{F}), $a_q=c_V^2\left[1 + 
\frac {1}{7}\left(\frac{c_A}{c_V}\right)^2\right]$ (see Table~1), and
$n_B=p_{F_q}^3/\pi^2$ is the baryon density ($n_0 = 0.16~{\rm
fm}^{-3}$ is the empirical nuclear matter equilibrium density). This
may be compared with the emissivities in normal quark matter 
from the
bremsstrahlung ($q + q \rightarrow q + q + \nu + \bar\nu$) 
and modified Urca ({\it e.g.,} $d + d \rightarrow d + u + e^- + \bar\nu$) 
processes~\cite{Iwas}
\begin{eqnarray}
\epsilon_q^{brems}  &\simeq & 3\times 10^{19}~ N_\nu T_9^8 
\left(\frac{n_B}{n_0}\right)^{1/3}~{\rm erg~cm^{-3}~s^{-1}} \,,\\ 
\epsilon_q^{MUrca}  &\simeq &  \alpha_c^2 \cos^2\theta_c~  
\epsilon_q^{brems} \,,
\label{epsmb}
\end{eqnarray}
where $\alpha_c$ is the strong interaction coupling and $\theta_C$ is
the Cabbibo angle.  Note that both of these processes are suppressed
by superfluidity by a factor of $\sim\exp(-2\Delta /k_BT)$.  Thus, as
$T$ falls below $T_c$, the PBF process dominates the cooling in quark
matter.

In baryonic matter, the $\nu\bar\nu$ emissivity from the PBF process
is ~\cite{Flo,Yaks}
\begin{eqnarray}
\epsilon_B^{\nu\bar\nu} 
&\cong &1.65 \times 10^{21}
N_\nu T_9^7 \left( \frac {n_B}{n_0} \right)^{1/3}
\left(\sum_i F_i a_i~Y_i^{1/3} \frac {m_i^*}{m_i}  \right)
 ~{\rm erg~cm^{-3}~s^{-1}} \,,
\label{epsqt}
\end{eqnarray}
where $a_i\cong c_{Vi}^2$ and $Y_i=n_i/n_B$ is the baryon concentration of
species $i$.  The values of the neutral current hyperon couplings in
Table~1, which were calculated in \cite{RPL} by considering both the
SU(3) singlet and octet contributions to the baryon current operators,
suggest that in matter containing hyperons, the PBF process can occur
with hyperons of all species.  The values quoted in Table~1 of
\cite{ykl}, where the $\Lambda$ and $\Sigma^0$ couplings were found to
vanish and the $\Xi^-$ coupling was very small, differ significantly
from those in Table~1 probably due to the neglect of the singlet
contributions in Ref.~\cite{ykl}.  It is intriguing that at the same 
baryon density in baryonic and quark matter, 
$\epsilon_B^{\nu\bar\nu}/\epsilon_q^{\nu\bar\nu}
\simeq ({\rm constant}) \cdot (n_B/n_0)^{-1/3}$ lies in the
range $2-5$ for the densities of interest in the long-term cooling
of solar mass compact stars.

The bremsstrahlung ($n + n \rightarrow n + n + \nu +\bar\nu$) and
modified Urca process ($n + n \rightarrow n + p + e^- + \bar\nu$)
emissivities are~\cite{FM,HPR} 
\begin{eqnarray}
\epsilon_B^{brems} &\cong & 
4.4 \times 10^{19} 
N_\nu T_9^8
\left(\frac {m_n^*}{m_n}\right)^4 
\left( \frac {n_B}{n_0} \right)^{1/3}~R_{nn}
 ~{\rm erg~cm^{-3}~s^{-1}} \,, \nonumber \\ 
\epsilon_B^{MUrca} &\cong & 
10^{22}\left(Y_e\frac{n_B}{n_0} \right)^{1/3}
 ~{\rm erg~cm^{-3}~s^{-1}} \,,  
\end{eqnarray}
where $R_{nn}$ is a density dependent reduction factor, that lies in
the range 0.3--0.15 for densities in the range $0.5-3~n_0$, recently
computed in Ref.~\cite{HPR} utilizing on-shell $nn$ amplitudes
(similar reductions are expected for the modified Urca process also). 

So far, we have restricted ourselves to the case of isotropic $^1{\rm
S}_0$ pairing, which provides a larger gap compared to anisotropic
p-wave pairing ($^3{\rm P}_2$ states).  The fact that the smallest gap
is the most relevant for long-term cooling highlights the importance
of p-wave pairing though it requires higher energies (or densities)
than s-wave pairing.  For example, for neutrons in the lowest energy
$m_{J} = 0$ quantum number of the $^3{\rm P}_2$ state, the anisotropy
of the gap yields the relation~\cite{Yaks}
\begin{equation}
\Delta(T) = 1.98k_BT_{c}(1-T/T_{c})^{1/2} \hspace{0.3in} (T_{c} - T \ll
T_{c}) \,.
\end{equation}

Just as for nucleons, the quark field operator in Eq.~(\ref{qfop})
must be modified to matrix-valued creation and annihilation operators
due to the triplet nature of pairing. It can be obtained by a unitary
transformation on the ground state~\cite{tam} with the coherence
factors in Eq.~(\ref{coh}) acquiring additional factors from the
angular dependence. The p-wave emissivity exhibits a $T$-dependence
that is similar to that from s-wave pairing.

The effects of strong and electromagnetic correlations on weak
interaction processes involving baryons are known to be
large~\cite{corrs}.  By considering only the electromagnetic
correlations between protons and electrons, Leinson has recently
demonstrated that $\nu\bar\nu$ emissivity from the PBF process from
gapped protons is substantially larger than that obtained at the mean
field level~\cite{lei}. A similar enhancement may be expected in quark
matter through a collective response of $u,d$, and $s$ quarks in a
charge neutral environment even with miniscule concentrations of
electrons.

Calculations to include effects germane to ``QCD condensed
matter''~\cite{CFL} would have to address modifications in the weak
interaction hamiltonian that will likely lead to couplings and
quasiparticle excitations that are modified from those considered
here.  We defer calculations of PBF processes in such phases and
detailed cooling simulations of stars with quark matter to
subsequent studies. \\

Research support from the DOE Grant No. FG02-88ER-40388 is gratefully
acknowledged.  We thank Dima Yakovlev for useful communications. \\

\newpage

\begin{table}
\begin{center}
\noindent {TABLE 1: The standard model neutral current vector and
axial--vector couplings of neutrinos to quarks and baryons.  The
hyperon couplings are taken from Ref.~\cite{RPL} and include both the
octet and singlet contributions.  $\ell$ stands for $e-, \mu-,$ or
$\tau-$ type neutrinos.  Numerical values are quoted using D=0.756,
F=0.477, and $\sin^2\theta_W$=0.23.}
\begin{tabular}{lcc}
{Reaction } & $c_V$  & $c_A $  \\
\hline
$\nu_\ell + u \rightarrow \nu_\ell + u $ & 
$\frac{1}{2}-\frac{4}{3}\sin^2{\theta_W}=0.1933$ & $\frac{1}{2}$ \\  
$\nu_\ell + d \rightarrow \nu_\ell + d $ & 
$\frac{1}{2}+\frac{2}{3}\sin^2{\theta_W}=0.6533$ & 
$-\frac{1}{2}$ \\  
$\nu_\ell + s \rightarrow \nu_\ell + s $ & 
$\frac{1}{2}+\frac{2}{3}\sin^2{\theta_W}=0.6533$ & $-\frac{1}{2}$ \\ \hline  
$\nu_\ell + n \rightarrow \nu_\ell + n$ & $-\frac 12$ & $-\frac 12(D+F)
=-0.6165$ \\
$\nu_\ell + p \rightarrow \nu_\ell+ p$ & $ \frac 12-2\sin^2\theta_W=0.04$ & 
$\frac 12(D+F)=0.6165$\\
$\nu_\ell + \Lambda\rightarrow \nu_\ell +\Lambda$ & $-\frac 12$ 
& $-\frac 12(F+\frac D3)=-0.3645 $ \\
$\nu_\ell + \Sigma^-\rightarrow \nu_\ell +\Sigma^-$ &
$ -\frac 32+2\sin^2\theta_W=-1.04$ & $\frac 12(D-3F)=-0.34$  \\
$\nu_\ell + \Sigma^+\rightarrow \nu_\ell +\Sigma^+$ &
$ \frac 12-2\sin^2\theta_W=0.04$ & $\frac 12(D+F)=0.6165$ \\
$\nu_\ell + \Sigma^0\rightarrow \nu_\ell +\Sigma^0$ & $-\frac 12$ 
& $\frac 12(D-F)=0.14$\\
$\nu_\ell + \Xi^-\rightarrow \nu_\ell +\Xi^-$ 
& $ -\frac 32+2\sin^2\theta_W=-1.04 $ & 
$\frac 12(D-3F)=-0.34$ \\
$\nu_\ell + \Xi^0\rightarrow \nu_\ell +\Xi^0$ & $-\frac 12$ 
& $-\frac 12(D+F)=-0.6165$ \\
\end{tabular}
\label{tab1}
\end{center}
\end{table}

\end{document}